\RequirePackage[loading]{tracefnt}
\documentclass[useAMS,usenatbib]{mn2e}

\usepackage{graphicx}
\usepackage{amssymb}
\usepackage{mathabx}
\usepackage{longtable,lscape}
\usepackage{natbib}
\usepackage{color}
\usepackage{ulem}

\def\ergscm{erg~s$^{-1}$~cm$^{-2}$}

\def\arcmin{\hbox{$^\prime$}}

\def\apj{ApJ}
\def\apjs{ApJ, Supplement}
\def\apjl{ApJ, Letters}

\def\aap{A\&A}
\def\mnras{MNRAS}
\def\nat{Nature}

\title[INTEGRAL 11-year hard X-ray survey above 100 keV]{INTEGRAL 11-year hard X-ray survey above 100 keV}
\author[Krivonos et al.]{R. Krivonos$^{1,2}$\thanks{E-mail:
krivonos@iki.rssi.ru (RK); stsygankov@gmail.com (ST)}, S. Tsygankov
$^{3,1}$\footnotemark[1], A. Lutovinov$^1$, M. Revnivtsev$^1$, \newauthor
E. Churazov$^{1,4}$ and R. Sunyaev$^{1,4}$\\
$^{1}$Space Research Institute, Russian Academy of Sciences, Profsoyuznaya 84/32, 117997 Moscow, Russia\\
$^{2}$Space Science Lab, University of California, Berkeley, CA 94720, USA;\\
$^3$Tuorla Observatory, Department of Physics and Astronomy, University of Turku, V\"ais\"al\"antie 20, FI-21500 Piikki\"o, Finland\\
$^4$Max Planck Institute for Astrophysics, Karl-Schwarzschild-Strasse 1, 85741 Garching, Germany}

\begin{document}

\date{\today}

\pagerange{\pageref{firstpage}--\pageref{lastpage}} \pubyear{2014}

\maketitle

\label{firstpage}

\begin{abstract}

We present results of all sky survey, performed with data acquired by the {\it IBIS} telescope onboard the INTEGRAL observatory
over eleven years of operation, at energies above 100 keV. A catalogue of detected sources includes
$132$ objects. The statistical sample detected on the time-averaged $100-150$~keV map at a
significance above $5\sigma$ contains $88$ sources: 28 AGNs, 38 LMXBs, 10 HMXBs and 12 rotation-powered young X-ray pulsars.
The catalogue includes also 15 persistent sources, which were registered
with the significance $4\sigma\leq S/N <5$ in hard X-rays, but at the same time were firmly detected ($\geq12\sigma$) in
the $17-60$~keV energy band. All sources from these two groups are known X-ray emitters, that means
that the catalogue has 100\% purity in respect to them. Additionally, 29 sources were found
in different time intervals.
In the context of the survey we present a hardness ratio of galactic and extragalactic sources,
a LMXBs longitudinal asymmetry and a number-flux relation for non-blazar AGNs.
At higher energies, in the $150-300$~keV energy band, 25 sources have been detected with a signal-to-noise ratio $S/N\geq5\sigma$,
including 7 AGNs, 13 LMXBs, 3 HMXBs, and 2 rotation-powered pulsars. Among LMXBs and HMXBs we identified
12 black hole candidates (BHC) and 4 neutron star (NS) binaries.

\end{abstract}


\section{Introduction}
A production of photons with energies above 100 keV requires either very high temperatures of the emitting
plasma or some non-thermal mechanisms, involving highly energetic/relativistic particles.
 Since it is not always clear what mechanism works, this energy range is under continuous investigations both from
theoretical and observational points of view.

Due to severe problems of a separation of hard X-ray photons from events, created by charged particles in hard
X-ray detectors, imaging capabilities of these instruments were very limited. First hard X-ray survey of the sky
was performed with scanning collimators of the {\it A4} instrument onboard the HEAO1 observatory \citep{levine84},
i.e. with the instrument, which was not optimized for the image reconstruction. The next step was done with the help of
the {\it SIGMA} coded mask telescope \citep{paul91} onboard the GRANAT observatory \citep{sunyaev90}, which
have provided the deepest (at that time) images of the Galactic Center region \citep{1991A&A...247L..29S,sunyaev91b} and
a large portion of the Galactic plane \citep{revnivtsev04} at energies 100-200 keV.

The latest generation of hard X-ray telescopes, like {\it IBIS} onboard the INTEGRAL observatory \citep{winkler03}
and {\it BAT} onboard the Swift observatory \citep{gehrels04}, which also used the coded mask imaging method,
have significantly larger effective areas and fields of view in a comparison with {\it SIGMA}. These advantages
together with much larger effective operational time allowed to produce hard X-ray surveys with a significantly
deeper sensitivity \citep{kri2012,baumgartner13}.

\begin{figure*}
  \centering
  \includegraphics[width=\textwidth]{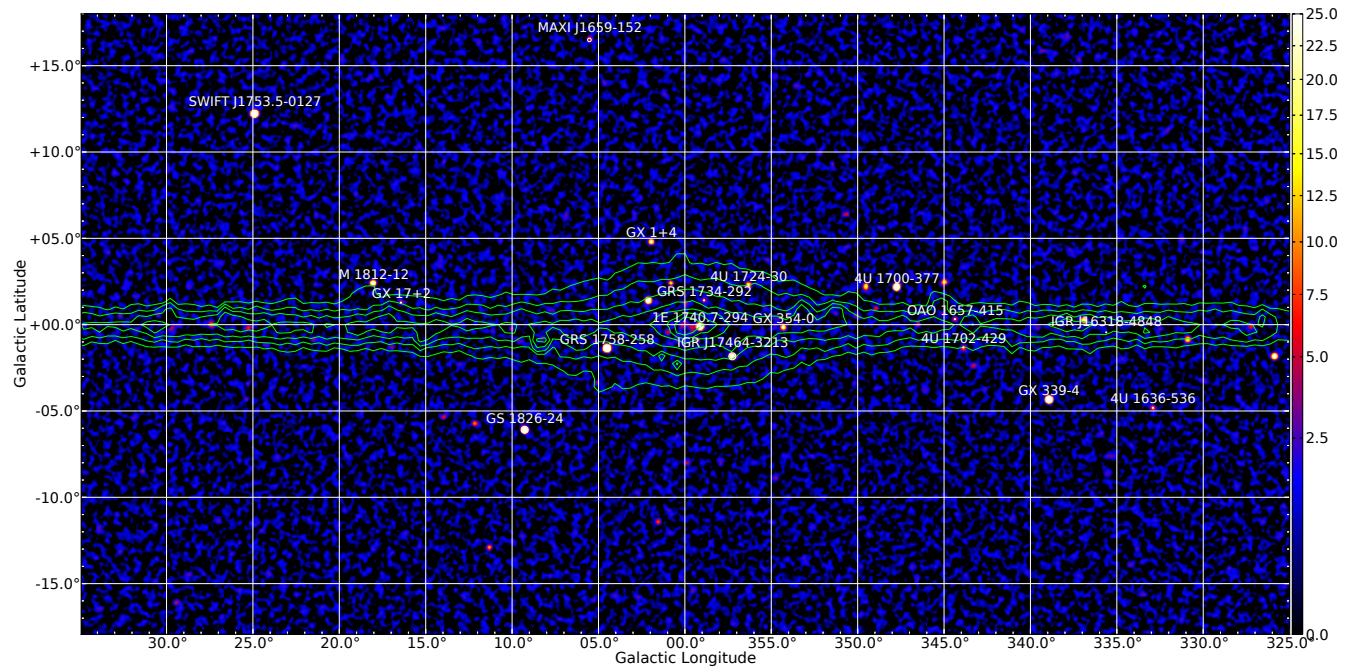}
  \caption{INTEGRAL/IBIS hard X-ray ($100-150$~keV) map of the Galactic bulge and (partially) disk.
    The total dead-time corrected exposure is about 30~Ms in the GC
    region. The image is shown in terms of $S/N$ with the color map
    ranging from values of 0 to 25. This color scheme is used to
    emphasize sky background variations. The green
contours are isophotes of the 4.9
micron surface brightness of the Galaxy (COBE/DIRBE) revealing its bulge/disk structure.}
  \label{fig:gcmap}
\end{figure*}

The {\it IBIS/INTEGRAL} telescope has demonstrated a great success in surveying of the hard X-ray sky at energies
above 20 keV, especially in sky areas with a high surface density of sources -- near the Galactic plane and Galactic Center.
While the best sensitivity of this instrument is achievable at lower energies (typically $\sim20-60$ keV), the
harder energy band $> 100-200$ keV might be potentially interesting due to study of a possible appearance of new
mechanisms of the photons generation. In this paper we present an 11-years
survey of the sky with the {\it IBIS} telescope onboard the INTEGRAL observatory at energies above 100 keV.
This survey is based on the substantial extension in the exposure with respect to previous INTEGRAL hard X-ray
surveys at these energies \citep[][hereinafter referred to as B06]{bazzano2006}.

\section{Data analysis}

For the current hard X-ray survey we utilized all publicly available INTEGRAL data acquired
 by the {\it IBIS} telescope \citep{ibis} between December 2002
and January 2014 (spacecraft revolutions 26-1377).
The survey also contains private data from M82 deep field (PI: Sazonov), and scanning observations of the Galactic Center
(PI: Krivonos) and Puppis region (PI: Tsygankov).  The coded-mask
telescope {\it IBIS} has a wide field of view $28^{\circ}\times28^{\circ}$
($9^{\circ}\times9^{\circ}$ fully coded) and moderate angular resolution of $12$\arcmin.
{\it IBIS} provides a localization accuracy of $<$2--3$\arcmin$ which is sufficiently enough
for the search of soft X-ray and optical counterparts and a subsequent optical
classification of newly discovered hard X-ray sources.
An every individual INTEGRAL observation, or so called
\textit{Science Window} (\textit{ScW}), with a typical exposure of $2$~ks was analyzed with a specially
developed software package (see e.g. \citealt{kri2010a,chur2014} and references therein) to produce sky
images in four energy bands: $17-60$, $60-100$, $100-150$, and $150-300$~keV.
Following our previous hard X-ray survey \citep{kri2012}, and to account for the
ongoing detector degradation and loss of the sensitivity at low energies, the flux scale in each \textit{ScW}
sky image was adjusted using the flux of the Crab Nebula taken from
the nearest observation.

After applying selection criteria over the list of reconstructed \textit{ScW}
sky images, as described in \cite{kri2007}, we obtained 95180 \textit{ScW}s in each band, which comprises
$\sim170$~Ms of the effective exposure. Individual sky
images were projected onto $25^{\circ}\times25^{\circ}$ sky frames covering
the whole sky in the \textit{HEALPix} reference grid \citep{healpix} with 192 tiles in total.
For the visualization purposes we accumulated sky mosaics in the Cartesian projection aligned with the Galactic
plane as it was done for the nine-year INTEGRAL Galactic hard X-ray survey\footnote{http://hea.iki.rssi.ru/integral} \citep{kri2012}.
Fig.~\ref{fig:gcmap} shows the Galactic bulge and disk map in the $100-150$~keV energy band.

\begin{figure}
\includegraphics[width=0.48\textwidth]{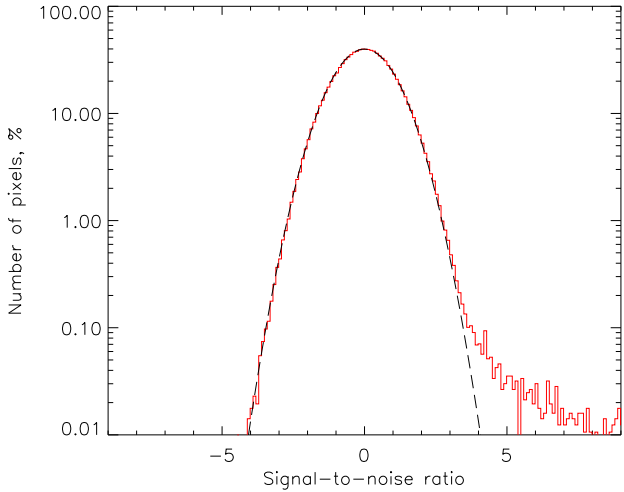}
\caption{Signal-to-noise ratio distribution of a number of pixels in
  the $100-150$~keV hard X-ray image shown in Fig.~\ref{fig:gcmap}. The dashed line
  represents the normal distribution with unit variance and zero
  mean.}\label{fig:gcmap:histo}
\end{figure}

{\sl It is important to emphasize here that the sensitivity of our survey is limited by a photon statistics
only and not affected by a systematic noise}, which severely limits the sensitivity achievable at lower energies
\citep[see, e.g.,][]{kri2010b,kri2012}.
This fact is demonstrated in Fig.~\ref{fig:gcmap:histo}, where we show a signal-to-noise
distribution of fluxes in pixels along with the Normal distribution representing a statistical noise.
The positive tail of the distribution is formed by real source counts.

\begin{figure}
 \includegraphics[width=0.48\textwidth]{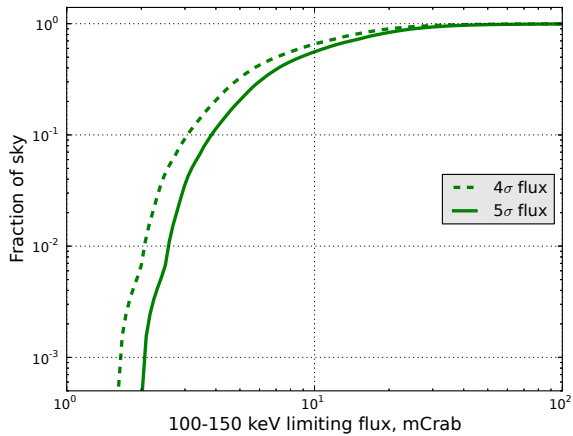}
\caption{Fraction of the sky covered as a function of $4\sigma$ (dashed line) and $5\sigma$ (solid line)
limited flux.}\label{fig:area}
\end{figure}

The survey sky coverage as a function of a $5\sigma$ limiting flux is shown in Fig.~\ref{fig:area}.
The peak sensitivity of the survey is about 2~mCrab ($8\times10^{-12}$ \ergscm\ in the $100-150$~keV energy band).
The survey covers $\simeq10$\% of the sky down to the flux limit of 3.7~mCrab ($1.5\times10^{-11}$ \ergscm)
and 90\% of the sky at 25~mCrab ($10^{-10}$ \ergscm).
To compare our sensitivity with the previous survey conducted in this energy band by B06,
we plotted also the covered sky area at a $4\sigma$ limiting flux, which was used by these authors.
As a result the current survey more then doubled the covered sky area at the limiting flux of 10~mCrab:
$\sim$65\% sky fraction versus $\sim$27\% in B06.

\subsection{The average sky}
\label{aver}

The source detection has been done on each mosaic image in the
reference $100-150$~keV energy band. The angular resolution of the {\it IBIS} telescope is  $12^\prime$,
which gives $\sim10^{6}$ independent pixels on the whole sky.
In order to ensure a false detection of not more than one source at the whole sky
we should adopt a $5\sigma$ detection threshold for the time-average sky if the distribution of signal-to-noise
ratios is purely statistical. To check it,  we divided the fluxes by errors, estimated for pure photon counting noise and
analyzed this distribution for our maps. An example of such a
distribution, obtained in the Galactic Center region is shown in Fig.\ref{fig:gcmap:histo}. Is it seen that this
distribution is very close to the pure Gaussian one. With this distribution the adopted $5\sigma$ detection
threshold indeed allows not more than one false detection over the whole sky. At the same time it can be seen also
from Fig.\ref{fig:gcmap:histo} that there exist a number of real excesses on the sky with a statistical significance
in the range of $4\sigma\leq S/N <5\sigma$.

In order to take into account (to include) such sources in the catalogue we applied a second detection criterion:
the source candidate, which is registered with a signal-to-noise ratio $4\sigma \leq S/N < 5\sigma$
and have a $12\sigma$ (or more) detection in the $17-60$~keV energy band is also included in the catalogue.
The adopted ratio of $S/N$ in the energy ranges $17-60$~keV and $100-150$~keV approximately corresponds
to the Crab-like spectrum of a source.

Using the first detection criterion  ($S/N \geq 5\sigma$) we have found 88 sources over the whole sky.
Among them there are 38 low-mass X-ray binaries (LMXBs), 10 high-mass X-ray binaries (HMXBs), 28
active galactic nuclei (AGNs), and 12 of other types.


An application of the second criterion (source candidates in the range $4\sigma \leq S/N < 5\sigma$ but with $S/N \geq 12\sigma$
in the $17-60$~keV energy band) led to the detection of 15 additional sources of a different nature: 7~LMXBs, one~HMXB,
6~AGNs, and one PWN.

It is important to note that, in contrast to the first (main) selection criterion,
the second one  has a more complicated selection function. Therefore
only the $5\sigma$ sample will be used below for statistical analysis, e.g.,
for studying the AGN surface density -- flux relation known as $Log$N$-Log$S distribution.

It is worth to mention as well that all detected hard X-ray ($100-150$~keV) sources
were also registered in the softer energy band $17-60$~keV. This means
that {\it we have not detected objects with exotic emission mechanisms and extremely hard spectra}.

\smallskip
{\noindent \sl SN~2014J in M82.}
 The only counter example (and only partial, because the source under the discussion
was not detected at level more than $4\sigma$ in the $100-150$~keV band) to this statement is the
transient emission from the supernova explosion SN~2014J in the galaxy M82.
This source was detected at a significance level of $\sim3.7\sigma$ in the energy
band $100-600$~keV with the highest flux at energies $200-300$~keV \citep{chur2014}
and remained undetected at energies below 100 keV. The extreme hardness of the emission of SN~2014J
is a result of unusual formation of its photon spectrum.
Usually photons at energies $100-300$~keV appear as a result of the Compton upscattering
of softer seed photons on hot electrons, while in the case of the supernova
explosion the hard X-ray photons diffuse from harder energy range $800-1200$~keV as a result
of the Compton downscattering on colder electrons, or produced via three-photon annihilation of positrons. Up to date only two sources with such a type
of spectrum were detected over all history of the X-ray astronomy -- supernova SN~1987A \citep{sunyaev87}
and supernova SN~2014J \citep{chur2014}. Another example of the Compton downscattering 
predicted at energies below 100 keV or so called ``Compton hump'' \citep{sunyaev1980}, has been recently observed in great detail with {\it NuSTAR} 
in some Compton-thick AGNs \citep{balokovich2014}.

\begin{table*}
\begin{minipage}{\textwidth}
\caption[Catalogue of hard X-ray sources detected in the $150-300$~keV band]
{Catalogue of hard X-ray sources detected in the $150-300$~keV band at $S/N\geq5\sigma$. Description of table columns see in Sect.~\ref{sect:catalog}} \label{tab:catalog:high}
 \begin{tabular}{@{}clrrccccl}
  \hline
\hline
No. & Name & Ra & Dec & Flux$_{\rm 150-300~keV}$ & Type & Ref.$^{1}$ & Notes$^2$ \\
    &      & deg & deg & $\textrm{erg~} \textrm{cm}^{-2}\textrm{s}^{-1}$ &   &  &  \\ \hline
1  &  {\bf                  Crab}  &  83.63  &  22.02  & $657.7 \pm 1.2$ &      PSR/PWN  &     &  TeV J0534+220; \\
2  &  {\bf              NGC 2110}  &  88.05  &  -7.46  & $13.7 \pm 2.3$ (5.9)     &  AGN  &     &   Sy2 z=0.007579; \\
3   &  {\bf              NGC 4151}  &  182.63  &  39.41  & $17.9 \pm 1.5$ &     AGN  &     &   Sy1 z=0.003262; \\
4   &  {\bf              NGC 4388}  &  186.45  &  12.66  & $6.3 \pm 1.1$ (5.9)    &  AGN  &     &   Sy2 z=0.008426; \\
5   &  {\bf                 3C273}  &  187.28  &  2.05  & $15.7 \pm 1.0$ &     AGN  &     &   Blazar; z=0.15834; \\
6   &  {\bf                 Cen A}  &  201.36  &  -43.02  & $47.0 \pm 2.0$ &     AGN  &     &   Sy2 z=0.001830; \\
7   &  {\bf           PSR B1509-58}  &  228.48  &  -59.14  & $13.3 \pm 1.5$ (8.8) &     PSR  &     &   \\
8    &   {\bf         XTE J1550-564}   &   237.76   &   -56.46   &  $14.0 \pm 1.3$  &      LMXB   & 1 &     BHC; \\
9     &    {\bf            4U 1630-47}    &    248.52    &    -47.39    &   $7.7 \pm 0.9$ (8.4)          &    LMXB    & 2 &     C; BHC; \\
10   &   {\bf              GX 339-4}   &   255.71   &   -48.79   &  $33.0 \pm 0.9$  &       LMXB   & 3 &    BHC; \\
11   &  {\bf           4U 1700-377}  &  255.98  &  -37.84  & $10.6 \pm 0.8$ &     HMXB  &     &   NS; \\
12    &    {\bf       IGR J17091-3624}    &    257.29    &    -36.41    &   $5.3 \pm 0.8$ (6.5)          &    LMXB    & 4,5,6 &     C; BHC; \\
13  &  {\bf              GX 354-0}  &  262.99  &  -33.83  & $4.7 \pm 0.7$ (6.6) &     LMXB  &             &   burster; \\
14    &    {\bf         1E 1740.7-294}    &    265.98    &    -29.73    &   $23.1 \pm 0.7$        &    LMXB    & 7 &     C; BHC; \\
15   &   {\bf Swift J174510.8-262411}   &   266.30   &   -26.40   &  $14.3 \pm 0.8$  &      LMXB   & 8 &    BHC; \\
16   &   {\bf       IGR J17464-3213}   &   266.56   &   -32.23   &  $10.0 \pm 0.7$  &       LMXB   & 9,10 &    H1743-322/XTE J1746-322; BHC; \\
17    &   {\bf    SWIFT J1753.5-0127}   &   268.37   &   -1.45   &  $63.4 \pm 1.5$  &        LMXB   & 11,12 &    BHC; \\
18   &   {\bf          GRS 1758-258}   &   270.30   &   -25.74   &  $40.8 \pm 0.7$  &       LMXB   & 13 &    BHC; \\
19    &   {\bf             M 1812-12}   &   273.78   &   -12.09   &  $8.6 \pm 1.0$ (8.7)       &   LMXB   & 14 &    burster; \\
20    &   {\bf            GS 1826-24}   &   277.37   &   -23.80   &  $12.5 \pm 0.9$        &   LMXB   & 15 &    burster; \\
21    &   {\bf          GRS 1915+105}   &   288.80   &   10.95   &  $32.9 \pm 1.0$        &   LMXB   & 16 &    BHC; \\
22   &   {\bf               Cyg X-1}   &   299.59   &   35.20   &  $416.1 \pm 1.1$      &   HMXB   & 17 &    BHC; \\
23   &  {\bf              Cygnus A}  &  299.87  &  40.74  & $7.2 \pm 1.2$ (6.2) &     AGN  &     &   Sy2 z=0.056146; =3C 405.0; \\
24   &   {\bf               Cyg X-3}   &   308.11   &   40.96   &  $11.2 \pm 1.1$  &        HMXB   & 18 &    BHC; \\
25   &   {\bf              3C 454.3}   &   343.49   &   16.15   &  $15.9 \pm 2.2$ (7.2)  &        AGN   & 19 &    Blazar; z=0.859; \\
\hline
 \end{tabular}

$^{1}$~References: (1) \cite{1998IAUC.7008....1S}, (2) \cite{1986ApJ...308..199P}, (3) \cite{1979Natur.278..434S},
(4) \cite{2003ATel..149....1K}, (5) \cite{2003AstL...29..719L}, (6) \cite{2006ApJ...643..376C},
(7) \cite{1991SvAL...17...54S}, (8) \cite{2012ATel.4381....1V}, (9) \cite{2003ATel..132....1R},
(10) \cite{2006ApJ...639..340K}, (11) \cite{2005ATel..550....1M}, (12) \cite{2014MNRAS.445.2424N},
(13) \cite{1991A&A...247L..29S}, (14) \cite{1983PASJ...35..531M}, (15) \cite{1999ApJ...514L..27U},
(16) \cite{2001A&A...373L..37G}, (17) \cite{1995A&A...297..556H}, (18) \cite{1996A&A...311L..25S}, (19) \cite{2005A&A...433.1163D}.

$^2$The spatial confusion with another source (mostly detected at energies below $100$~keV) is indicated by sign C.
Names of sources in the confusion can be found in the full online version of Table~\ref{tab:catalog:short}.
The measured flux of sources being in the spatial confusion should be taken with the caution.
\end{minipage}
\end{table*}

We also investigated a sky map in the harder energy band $150-300$~keV and found 25 significantly ($S/N \geq 5\sigma$)
detected sources of a different nature: 7~AGNs (one Seyfert type~1, 4 Seyfert type~2 galaxies, 2~blazars
at high redshifts), 13~LMXBs, 3~HMXBs (Cyg X-1, Cyg~X-3 and 4U~1700-377), and 2~rotation-powered pulsars (PSRs: Crab and PSR~B1509-58).
Among LMXBs and HMXBs we identified 12 black hole candidates, the neutron star binary system 4U~1700-377 and 3 X-ray
bursters (GX~354-0, M~1812-12 and GS~1826-24).
All these sources have been detected in the $100-150$~keV energy band as well
and therefore present in the catalogue. Table~\ref{tab:catalog:high} lists all the sources detected in the $150-300$~keV band
with corresponding fluxes and source types.

\begin{figure*}
  \centering
  \includegraphics[width=\textwidth,bb=-19 230 623 555,clip]{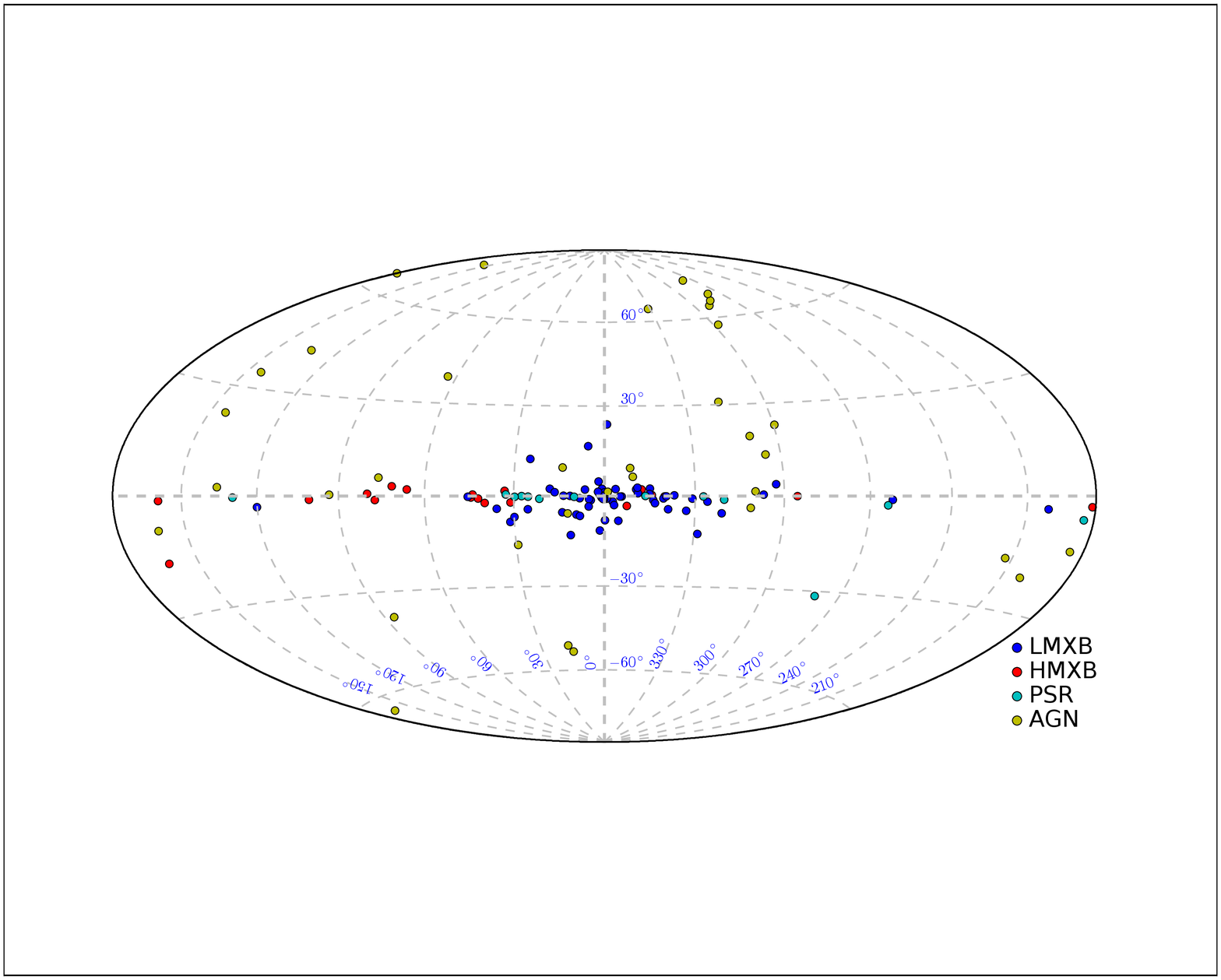}
  \caption{All-sky map showing three basic types of X-ray sources detected in the $100-150$~keV survey:
64 LMXBs, 19 HMXBs and 35 AGNs, encoded by blue, red and yellow colors, respectively.
The map was produced according to the Hammer-Aitoff projection in galactic coordinates.}
  \label{fig:aitmap}
\end{figure*}

\subsection{Timing analysis}
\label{timing}

The sky averaged over the time period of eleven years is optimized for the detection of persistent
sources, therefore, transient events or periods of a sources activity can be missed in such a survey.

In order to recover periods of the activity of transients, we measured the flux of known sources, detected in our
previous surveys ($\sim800$ sources, see \citealt{kri2007,kri2010b,kri2012}),
over each spacecraft revolution (3 days). Further these data were combined into a lightcurve in order
to estimate the source $S/N$ ratio in sliding time windows and to find periods of its significant excess.

More precisely, we averaged the source flux over all orbits in the range of $R_i\pm R_{w}$, where $R_i$ is the
current revolution and a half-width of the time window $R_w$ is varied from 0 to 400 orbits. It allow to cover time
scales from 3 days (one orbit) to $\sim6.7$ years, which is nearly half-time of the whole time span of the survey (1380 orbits or 11 years).
For the source detection we applied the second detection criterion, requiring the $S/N\geq 4\sigma$ detection in the
$100-150$~keV energy band and the $S/N\geq 12\sigma$ soft $17-60$~keV counterpart.

As a result, we found periods of the activity for 29 catalogued sources, which were not detected on the
time-average map. Taking four objects with tentative classification into account, the list of transients includes 20~LMXBs, 8~HMXBs,  and AGN Seyfert type~2 MRK 3, \citep{mrk3}. Thus, galactic sources (LMXBs and HMXBs) are the dominant among transients at energies above 100~keV.

It is interesting to note, that a number of transient HMXBs is comparable with a number of
persistent sources of this class, detected above 100 keV -- 8 vs 11, or $72$\%. At the same time
a relative part of transient LMXBs in this energy band is significantly lower -- 20 vs 45, or
$44$\%. This may be due to the fact that a vast majority of transient HMXBs are X-ray pulsars
(presumably in binary systems with Be stars). From time to time such objects show a strong
outburst activity and despite their relatively soft spectrum  become bright enough to be detected above 100 keV.
Sometimes such sources demonstrate an additional hard X-ray component in their spectrum as well
(see, e.g., the X-ray pulsar RX\,J0440.9+4431, \citealt{tsy2012}).




\section{Discussion}

The sky at energies $100-150$~keV at our flux cut of a few mCrab is dominated by galactic sources (60 out of 88),
including 38 LMXBs, 10 HMXBs and 12 PSRs.
Compared to the INTEGRAL Galactic nine-year $17-60$~keV survey at $|b|<17.5^{\circ}$ \citep{kri2012},
the number of LMXBs and HMXBs drops by a factor of $\sim3$ and $\sim8$ correspondingly.
Such a dramatic difference is indeed expected as HMXBs have softer spectra than LMXBs.
Indeed, the vast majority of high mass X-ray binaries, seen by INTEGRAL, are magnetic neutron stars,
accreting matter from their binary companions \citep{liu2006,lut2009,lut2013}.
The emergent spectra of these sources are generated by a hot plasma, heated in an accretion
column near the neutron star surface \cite[e.g.][]{davidson73,basko76,nagel81,becker05},
and usually demonstrate an exponential cutoff at high energies \citep[see, e.g.,][]{whi83,fil05}.
In several cases a hard X-ray component was found in spectra of HMXBs
\citep[see, e.g.,][]{disalvo98,bar08,dor12,lut2012}, that allows one to detect them above 100 keV
in the persistent state.

An emission of low mass X-ray binaries is formed in the innermost regions of the
accretion flow around typically non-magnetic sources. Extensive studies of these sources show
that they demonstrate different spectral states \cite[e.g.][]{esin97,poutanen97,barret00,disalvo00}. Typically LMXBs have hard spectra (both BH and NS) during low level of their mass accretion rate and soft spectra at high mass accretion rates. The total number of LMXBs in INTEGRAL Galactic nine-year survey is mainly provided by faint sources (i.e. sources with low level of mass accretion rate). It means that their spectra are hard (harder than those of HMXBs) and thus we should expect that the ratio of number of LMXBs in 17-60 keV and 100-150 keV should be smaller than that for HMXBs.

Another class of galactic sources detected above 100~keV are rotation-powered young X-ray pulsars. They typically demonstrate hard
power-law spectra with photon indexes of 1.5-2.0 at energies above 20 keV. It is widely accepted that 
their hard X-ray emission is dominated by the extended pulsar wind nebula (PWN) emitting synchrotron photons 
and Compton upscattering of softer photons on relativistic electrons \cite[e.g.][]{harding05}. The INTEGRAL nine-year 
Galactic plane survey in the $17-60$ keV energy band contains 16 PSRs at $|b|<17.5^{\circ}$ \citep{kri2012}. 
The current $100-150$~keV survey includes 13 PSRs, one of which --  PSR~0540-69 -- the pulsar and 
supernova remnant in the Large Magellanic Cloud. As a result only 4 galactic ($|b|<17.5^{\circ}$) PSRs 
detected in the $17-60$~keV energy band are below of the detection threshold in the $100-150$~keV energy band. 
Thus PSRs becomes a second dominant class of galactic hard X-ray sources (after LMXBs) above 100 keV. This clearly 
demonstrates that a non-thermal emission mechanisms start to play an important role at these energies.

\subsection{Hardness ratio}
\label{hardness}

\begin{figure}
\includegraphics[width=\columnwidth]{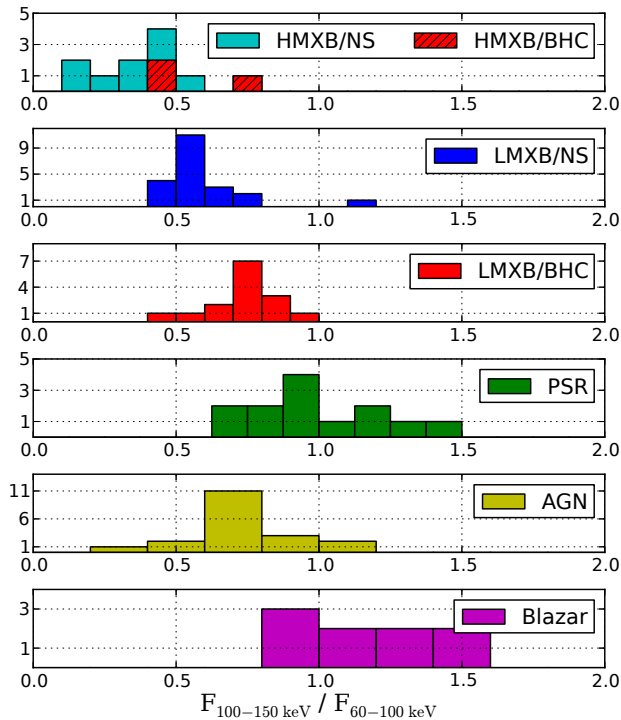}
\caption{Histogram of X-ray hardness ($100-150$~keV/$60-100$~keV) for different types of sources.
The width of bins has been adjusted in accordance with an uncertainty of hardness measurements.
}\label{fig:plot:hr2:hist}
\end{figure}

A hardness ratio characterizes a steepness of the source spectrum and can be calculated as a
ratio of a source flux in two adjacent energy bands. Assuming Crab-like spectra
we converted $60-100$ and $100-150$~keV fluxes from units of mCrab to \ergscm\ and calculated their ratio.
Hardness ratio histograms of different classes of sources detected with using of the first condition
($S/N \geq 5\sigma$) are shown in Fig.~\ref{fig:plot:hr2:hist}.

Several facts can be easily seen from this figure. In particular, BHCs tend to have harder power-law slopes,
than NSs in low mass X-ray binaries, which demonstrate softer spectra; PSRs have even harder spectra; non-blazar AGNs occupy a similar
region to that of LMXB/BHCs; blazars are shifted in the harder domain.
Hardness of HMXBs is below than that of LMXBs and AGNs, which reflects the known fact that
spectra of HMXBs are softer in general (see above for details). Three known HMXB/BHCs --
Cyg~X-1, Cyg X-3 and SS433 are also highlighted in Fig.\ref{fig:plot:hr2:hist}.
 Thus, we can conclude that Fig.~\ref{fig:plot:hr2:hist}
demonstrates a gradual hardening of spectra from HMXBs through LMXBs to AGNs,  and from NS to BHC.
 Of course one should bear in mind that all these objects are selected based on their $100-150$~keV flux.

\subsection{LMXBs longitude asymmetry}

\cite{wieden2008} reported a distinct asymmetry in the 511-keV line
emission coming from the inner Galactic disk. Authors argued that this asymmetry resembles
an asymmetry in the distribution of LMXBs, indicating that they
may be a dominant origin of positrons (a hypothesis that subclass of LMXBs --
so-called microquasars -- can be responsible for the positron production
was proposed, e.g., by \citealt{guessoum06}, and verified observationally by \citealt{2010AstL...36..237T}).
Based on the INTEGRAL hard X-ray survey above $20$~keV,
performed by \cite{bird2007}, \cite{wieden2008} demonstrated that a number
of LMXBs at negative longitudes (45) is higher than that at positive longitudes
(26) by a factor of $\simeq1.7$, and at higher energies the ratio becomes even larger (B06).
We investigated this LMXB asymmetry using greatly improved high-completeness
hard X-ray surveys by \cite{kri2012} and this work.

The 9-year Galactic survey \citep{kri2012} contains 108 LMXBs detected in the 17-60 keV band
at $S/N\geq4.7\sigma$, 37 of them are found at positive and 54 at negative longitutes
(hereafter within the selection box $|l|<50^{\circ}$ and $|b|<10^{\circ}$), comprising
the ratio $\simeq1.5$. Despite the apparent excess of LMXBs right to the Galactic center (GC),
the presumption against null hypothesis or symmetric distribution is rather low ($p=9.3\%$).

The number of LMXBs detected with $S/N\geq5\sigma$ on the average $100-150$~keV map of the
current survey at negative longitudes (19) is higher than that at positive longitudes
(11) by a factor of $\simeq1.7$  (chance probability $p=20\%$).
The corresponding flux-weighted ratio $\sim1.8$ is lower than
$2.8$ estimated by \cite{wieden2008} from the previous $100-150$~keV survey (B06).
LMXBs designated as black hole candidates demonstrate a similar
asymmetry $8:5$ at low significance ($p=58\%$).

Compared LMXB counts on both sides from GC in $17-60$ and $100-150$~keV bands, we confirm LMXB asymmetry
derived from B06, but do not see significant growth of the ratio with energy.

\section{Number-flux relation (LogN--LogS)}

{\it Extragalactic population.} The sample of sources detected at $S/N\geq5\sigma$ in the $100-150$~keV energy range on
time-averaged sky maps contains 29 extragalactic sources. Apart from the young pulsar PSR 0540-69
in Large Magellanic Cloud all other sources are AGNs. We removed nine blazars from this AGN sample (Mrk\,421,
MKN\,501, IGR\,J16562-3301, 3C\,454.3, PKS~2149-306, PKS~1830-211, 3C279, PKS~1219+04 and S5~0836+71)
to construct a non-blazar AGN selection and to measure their surface
density over the sky. Note, that our 19 AGN sample doubles the $100-150$~keV AGN
selection by B06, which contained 10 sources.

\begin{figure}
\includegraphics[width=\columnwidth]{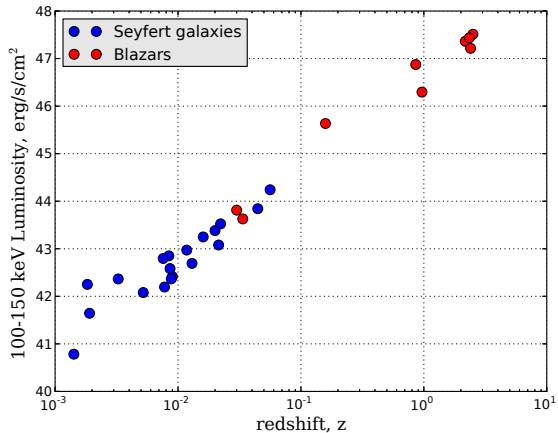}
\caption{Redshift -- luminosity diagramm for 28 Seyfert galaxies and blazars detected over the whole sky at $S/N>5\sigma$ in $100-150$~keV.}\label{fig:plot:lum}
\end{figure}

\begin{figure}
\includegraphics[width=\columnwidth]{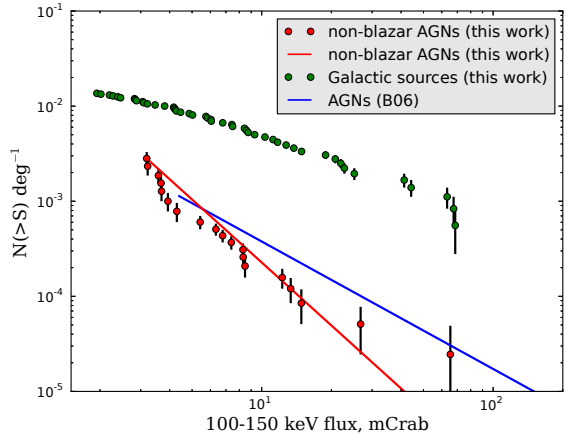}
\caption{Surface density of non-blazar AGNs (red points) and galactic sources (green points) as a function
of flux in the $100-150$~keV energy band. Galactic sources are counted only at $|b|<5^{\circ}$. No correction
for the survey sensitivity is done for them.
Lines represent AGNs number-flux relations from this work (red) and from B06 (blue).}
\label{fig:plot:lognlogs}
\end{figure}

Fig.~\ref{fig:plot:lum} shows a luminosity versus redshift diagram of the selected AGN sample.
 It demonstrates the fact that our AGN sample is dominated by nearby AGNs ($z<0.02$).
At the same time, we can detect very distant
AGNs (mostly blazars) generating a collimated emission in the high energy domain.
Some of them, namely blazar 3C273 (z=0.158), have been even detected in the $150-300$~keV energy
band (see Table~\ref{tab:catalog:high}).
The sensitivity-corrected cumulative $\log N-\log S$ relation for AGNs from our sample is shown in Fig.~\ref{fig:plot:lognlogs}.
This $\log N$--$\log S$ distribution can be adequately fitted by a power law $N(>S)=A
S^{-\alpha}$. Using a maximum-likelihood estimator \citep[see, e.g.,][]{jauncey67,crawford1970},
we determined the best-fit values of the slope and normalization: $\alpha=2.2\pm0.4$ and
$A=(2.27\pm0.52)\times10^{-4}$ deg$^{-2}$ at $S=10$~mCrab ($4.0\times10^{-11}$
\ergscm). The observed $\log N$--$\log S$ shape doesn not look smooth and its slope is somewhat steeper than expected for
the homogeneous distribution of sources in space ($\alpha= 3/2$),  which is probably caused by (i) deep field observations around known AGNs and/or (ii) deep field observations of the sky regions where nearby large scale structure can affect number-flux relation. In support of this assumption one can point out to the deep observations around 3C~273 and the Coma Cluster \citep{kri2005,paltani2008}, which are located in the region of the highest mass concentrations in the local Universe, well traced in hard X-rays by AGN spatial distribution \citep{kri2007,majello2012}. \cite{kri2007} demonstrated that nearby mass inhomogeneity can strongly affect AGN number-flux relation. Our $\log N$--$\log S$ is also steeper than that measured by B06 ($\alpha=1.34\pm0.04$). The latter is also shown in Fig.~\ref{fig:plot:lognlogs} and plotted down to the flux of the weakest source in their sample (4.4~mCrab, PKS~1830-211, however removed in our AGN sample). As seen from the figure we extended the hard X-ray ($100-150$~keV) $\log N$--$\log S$ down to the flux of 3.2~mCrab (NGC~4235) or by a factor of $\sim1.4$.

AGNs with fluxes exceeding the survey $5\sigma$ detection threshold account for $\sim1\%$
of the intensity of the cosmic X-ray background (CXB) in the $100-150$~keV band,
based on the CXB spectrum of \cite{gruber1999}.

{\it Galactic population.}
We constructed also a number-flux relation for galactic sources (Fig.~\ref{fig:plot:lognlogs}).
Similar to \cite{kri2007}, we selected sources at $|b|<5^{\circ}$, ending up with 49 objects, including 30 LMXBs, 9~HMXBs,
and 10 of other types. Since the population of galactic sources is not isotropically distributed in space
we cannot correct the $\log N-\log S$ diagram for the sensitivity map in the same way as we did it for AGNs.
Therefore the cumulative histogram shown in Fig. \ref{fig:plot:lognlogs} is just a number of sources divided
by the sky area within $|b|<5^{\circ}$. This number-flux of Galactic sources is useful for easy estimates of expected number of hard X-ray sources for future surveys of the Galactic plane.

\section{Catalogue of sources}
\label{sect:catalog}

The catalogue has been compiled from sources
passed through detection conditions in the reference $100-150$~keV energy band on time-average sky maps (Sect.~\ref{aver})
and maps built over different time periods (Sect.~\ref{timing}). A sample catalogue is shown in Table~\ref{tab:catalog:short}.
The text below describes columns of the catalogue table, which is also can be applied to Table~\ref{tab:catalog:high}
of sources detected in the $150-300$~keV energy band.

\begin{table*}
\begin{minipage}{\textwidth}
\caption[A sample from the catalogue of hard X-ray sources detected in the $100-150$~keV energy band]
{A sample from the catalogue of hard X-ray sources detected in the $100-150$~keV energy band.
The full version is available in the online version of this article.} \label{tab:catalog:short}
 \begin{tabular}{@{}clrrccccl}
  \hline
\hline
No. & Name & Ra & Dec & Flux$_{\rm 100-150~keV}$ & Range & Type & Ref.$^{1}$ & Notes \\
    &      & deg & deg & $\textrm{erg~} \textrm{cm}^{-2}\textrm{s}^{-1}$ & orbits &  &  &  \\ \hline
1  &   IGR J00291+5934  &  7.26  &  59.57  &  \textcolor{red}{$0.5 \pm 0.2$} (2.7)  &           &  LMXB  & 1,2 &   accreting millisecond pulsar; \\
-- &   &  &  & $11.5 \pm 0.9$  & 0261 0263 &  &  & V* V1037 Cas; \\
2 &  {\bf        PSR J0146+6145 } & 26.59 & 61.75 & $3.5 \pm 0.2$  &  & PSR &      &  AXP; 4U 0142+61; \\
3 &  {\bf               NGC 788} & 30.28 & -6.82 & $1.5 \pm 0.4$ (4.1) &  & AGN &         &  Sy2 z=0.0136; \\
4 &  {\bf       4U 0241+61} & 41.24 & 62.46 & $1.7 \pm 0.3$ (5.1) &  & AGN &         &  Sy1 z=0.044557; \\
5 &  {\bf      4U 0352+30} & 58.85 & 31.04 & $5.6 \pm 0.4$  &  & HMXB &     &  X Per; \\
6 &  {\bf                 3C111} & 64.58 & 38.02 & $1.6 \pm 0.4$ (4.0) &  & AGN &         &  Sy1 z=0.0485; \\
7  &   RX J0440.9+4431  &  70.23  &  44.56  &  \textcolor{red}{ $<$0.7 }  &           &  HMXB  & 3 &   \\
-- &   &  &  & $7.5 \pm 1.5$ (5.0) & 0963 0965 &  &  &  \\
 $<\dots>$  &  & & &  &  & & &  \\
132  &   {\bf 3C 454.3}  &  343.49  &  16.15  &  $5.0 \pm 0.4$   &    &  AGN  & 4 &   Blazar; z=0.859001; \\
\hline
 \end{tabular}

$^{1}$~References: (1) \cite{2004ATel..352....1E}, (2) \cite{2004ATel..353....1M}, (3) \cite{1999MNRAS.306..100R}, (4) \cite{2005A&A...433.1163D}.
\end{minipage}
\end{table*}

{\it Column (1) ``Id''} -- source sequence number in the catalogue.

{\it Column (2) ``Name''} -- source name.
According to our previous hard X-ray surveys \citep{kri2007,kri2010a,kri2012},
their common names are given for sources whose nature was known before their detection by
INTEGRAL. Sources discovered by INTEGRAL or whose nature was
established thanks to INTEGRAL observations are named ``IGR''.
The source name is highlighted by bold font if the source is significantly detected
on the time-averaged map.

{\it Columns (3,4) ``RA, Dec''} -- source equatorial (J2000) coordinates.
The positional accuracy of sources detected by {\it IBIS} depends on the source significance \citep{gros2003}.
The estimated $68\%$ confidence intervals for sources detected at 5--6, 10, and $>20\sigma$ are $2.1$\arcmin,
$1.5$\arcmin, and $<0.8$\arcmin,  respectively \citep{kri2007}.

{\it Column (5) ``Flux''} -- time-averaged source flux in the $100-150$~keV energy band.
Other energy bands -- $17-60$, $60-100$, and $150-300$~keV are available only online.
The flux measured at $S/N<4\sigma$ is highlighted in red. The flux with $S/N<2\sigma$ is shown as an $2\sigma$ upper limit, also in red color. If
the detection significance does not exceed $10\sigma$, it is shown in braces.
The flux is expressed in units of $10^{-11}\textrm{erg}$~$\textrm{cm}^{-2}\textrm{s}^{-1}$.

{\it Column (6) ``Range''} -- time interval in units of spacecraft orbits, when the corresponding flux
has been measured. If this column contains empty value, the flux has been determined from the whole
time span of the survey. INTEGRAL orbital revolutions can be roughly converted into the Modified Julian
Date (MJD) with the following empirical expression:
$55551.61223380 + ({\rm ORBIT}-1000.0)/0.33404581$.

{\it Column (8) ``Type''} -- general astrophysical type of the object:
LMXB (HMXB) -- low- (high-) mass X-ray binary, AGN -- active galactic
nucleus, SNR -- supernova remnant, PSR -- isolated pulsar or pulsar wind nebula (PWN),
SGR -- soft gamma repeater. A type in blue with a question mark indicates
that the specified type is not firmly determined and should be confirmed.


\section{Summary}

In this paper we have presented the catalogue of hard X-ray sources
detected in the $100-150$~keV all-sky survey performed by the INTEGRAL observatory
over 11 years of operation.
The current survey significantly improves our knowledge about a high energy sky compared to
previous survey conducted by \citet{bazzano2006}.
Our catalogue contains sources detected both on time-averaged sky maps -- optimized mostly
for persistent sources, and in different time intervals -- preferential for transients.
The survey has 100\% purity for persistent sources, while the list of transients contains
objects with tentative classification (4). Summary of source
types over different detection conditions is presented in Table~\ref{tab:summary}.

\begin{table}
\begin{center}
  \begin{tabular}{rccccll}
           & $S/N \geq 5\sigma$ & $4\sigma \leq S/N < 5\sigma$ & Time & Total \\ \hline
      LMXB & 38 & 7 & 20 & 65 \\
      HMXB & 10 & 1 & 8 & 19 \\
       PSR & 12 & 1 & 0 & 13 \\
       AGN & 28 & 6  & 1 & 35 \\ \hline
     Total & 88 & 15 & 29 & 132 \\
    \end{tabular}
  \caption{Number of different source types detected on time-averaged sky and different time intervals (``Time'').}
  \label{tab:summary}
\end{center}
\end{table}

The survey is dominated by 97 hard X-ray sources of the galactic origin (mainly, LMXBs and HMXBs -- in total 83),
in comparison with the extragalactic source population represented by 35 AGNs.

We provided hardness ratio ($100-150$~keV/$60-100$~keV) histograms for different source classes, showing a
gradual hardening of spectra from HMXBs and LMXBs through PSRs to AGNs, and from NS to BHC (Sect.~\ref{hardness}).

We characterized the $100-150$~keV non-blazar AGN population, collected from the survey, by the redshift-luminosity
diagram and number-flux relation based on 19 AGNs. Obtained results is compared to that published by \citet{bazzano2006}
and show that our $\log N$--$\log S$ is extended down to fainter fluxes by a factor of 1.4 and have a steeper slope.

The hard X-ray flux from the non-blazar AGN sample at $S/N\geq5\sigma$ accounts for $\sim1\%$ of the CXB intensity in the
$100-150$~keV energy band.


\section*{Acknowledgments}

Research is based on observations with INTEGRAL, an ESA project with instruments and science data centre funded by ESA member states (especially the PI countries: Denmark, France, Germany, Italy, Switzerland, Spain), Czech Republic and Poland, and with the participation of Russia and the USA. Research has made use of data obtained through the High Energy Astrophysics Science Archive Research Center Online Service, provided by the NASA/Goddard Space Flight Center. Authors thank Max Planck Institute fuer Astrophysik for computational support. The work was supported by grant of Russian Science Foundation 14-22-00271.


\label{lastpage}

\end{document}